# Hamiltonian-Generated Dynamics and Quantum Resource Allocation in a 4-Qubit Isotropic Heisenberg Spin Ring with Adjustable Next-Nearest-Neighbor Interactions

Seyed Mohsen Moosavi Khansari[1]
*Department of Physics, Faculty of Basic Sciences, Ayatollah Boroujerdi University, Boroujerd, IRAN*

**Abstract**
This investigation delivers a fully analytical solution for the unitary evolution of a four-qubit isotropic Heisenberg $XXX$ ring with periodic boundary conditions and a tunable next-nearest-neighbor exchange parameter $\alpha$. Adopting a Bell-type product state as the initial condition, we obtain closed-form expressions for the square-root fidelity $F(\rho(0),\rho(t)) = |\cos(\phi/2)|$, the global $\ell_1$-norm coherence $C_{\ell_1}(\rho(t)) = \sin^2(\phi/2)$, and the bipartite entanglement of formation $E_F(t)$ for the reduced pairs (1,2) and (3,4). These quantities are dictated entirely by the single phase $\phi = (\alpha+1)t$. The fidelity shows oscillatory behavior with amplitude $|\cos(\phi/2)|$ and reaches a quiescent dynamical regime at $\alpha = -1$, characterized by $F \equiv 1$ for all times. Meanwhile, the coherence follows $C_{\ell_1}(\rho(t)) = \sin^2(\phi/2)$, with a response sensitivity proportional to $|\alpha+1|$, and fully extinguishes at the stabilization point $\alpha = -1$. The entanglement of formation $E_F(t)$, formulated as an entropic function of $\phi$, undergoes striped periodic modulations and likewise freezes at the same parameter value. The unification of the phase ties together all observables, showing that larger $|\alpha+1|$ drives faster evolution, with peak susceptibility occurring when $(\alpha+1)t = \pi/4 + k\pi/2$. This complete analytical architecture supplies precise reference data suited for small-scale quantum hardware and charts a clear route for incorporating decoherence, thermal effects, and generalizations to larger spin assemblies.

## 1. Introduction

This study explores the unitary evolution and the concomitant quantum resource characteristics within a finite four-qubit isotropic Heisenberg $XXX$ framework, which includes modifiable next-nearest-neighbor couplings. The capacity to precisely steer quantum coherence and entanglement in few-qubit devices is an essential prerequisite for probing non-equilibrium many-body effects and for the advancement of near-future quantum technologies [1-4]. Although comprehensive computational analyses of larger arrays have yielded significant knowledge, there remains a notable divide between computationally intensive numerical techniques and experimentally viable few-qubit architectures that are suitable for a rigorous analytical treatment.

The current article addresses this shortfall by scrutinizing a minimal four-qubit Heisenberg $XXX$ ring with nearest-neighbor exchange (fixed at unity strength), supplemented by next-nearest-neighbor couplings governed by an adjustable parameter $\alpha$. The architecture constitutes a closed ring with periodic boundary conditions, thus allowing qubit N to couple with qubit 1.

The Hamiltonian (with $\hbar = 1$) for this periodic configuration [see Eq. (1)] takes the form:

$$H = \sum_{i=1}^{4} \frac{1}{4}\left(\sigma_i^X\sigma_{i+1}^X + \sigma_i^Y\sigma_{i+1}^Y + \sigma_i^Z\sigma_{i+1}^Z\right) + \sum_{i=1}^{4} \frac{1}{4}\alpha\left(\sigma_i^X\sigma_{i+2}^X + \sigma_i^Y\sigma_{i+2}^Y + \sigma_i^Z\sigma_{i+2}^Z\right)$$

[1] M.Moosavikhansari@abru.ac.ir

with periodic identification: $\sigma_5 \equiv \sigma_1$ and $\sigma_6 \equiv \sigma_2$

The system is prepared in a factorized arrangement of Bell-type character, namely $|\psi(0)\rangle = |\Psi^+_{1,2}\rangle \otimes |0_3, 0_4\rangle$, which, when expressed in the computational basis as given in Eq. (4), takes the specific form:

$$|\psi(0)\rangle = \frac{1}{\sqrt{2}}(|1_1, 0_2, 0_3, 0_4\rangle + |0_1, 1_2, 0_3, 0_4\rangle)$$

The central objective of this investigation is to obtain closed-form analytical solutions for the unitary time evolution and to examine the function of the parameter $\alpha$ in regulating three principal quantum resources:

1. Fidelity with respect to the initial configuration.
2. $\ell_1$-norm quantum coherence of the global four-qubit state.
3. Entanglement of formation for the two-qubit reduced subsystems (specifically partitions 12 and 34).

A central conceptual advance consists of identifying a unified phase variable $\phi = (\alpha + 1)t$ that encapsulates the complete dynamical behavior: all observable quantities depend solely on $\phi$ rather than on $\alpha$ and $t$ independently. This unification reveals periodic banded structures in the $(\alpha, t)$ plane with straight level sets along constant $\phi$ lines.

Particularly significant from an applied perspective is the identification of a stationary dynamical phase at the specific value $\alpha = -1$, wherein the phase variable $\phi$ remains identically zero across the entire temporal evolution, thus leading to static fidelity, fully suppressed coherence, and invariant entanglement. This fixed point serves as a straightforward tuning handle for safeguarding quantum informational assets in finite-size spin systems. In contrast, deviating $\alpha$ from this critical value systematically modifies the oscillatory frequency and parametric sensitivity, thereby facilitating the swift creation and dynamic regulation of both coherence and entanglement.

Although a wealth of numerical and thermodynamic studies on entanglement within spin chains are available [20, 25-28], rigorous closed-form analytical descriptions that simultaneously capture the dynamics of fidelity, coherence, and entanglement of formation in finite Heisenberg chains featuring adjustable next-nearest-neighbor ($NNN$) interactions remain comparatively uncommon. Prior efforts have examined analogous frameworks through Bethe ansatz techniques [27], exact numerical diagonalization [28], or quench dynamics [26]; nevertheless, none have successfully extracted the comprehensive phase architecture and stabilization mechanism that we present in this study. The present work fills this lacuna by delivering fully analytical expressions for a four-qubit setup, which uncover a unifying phase variable, $\phi = (\alpha + 1)t$, that dictates all examined quantum resources, and simultaneously pinpoint a robust stabilization scheme situated at $\alpha = -1$. Practical experimental realizations of this minimal model are achievable across numerous modern platforms, encompassing superconducting transmon qubits equipped with tunable couplers [22], trapped-ion arrays possessing adjustable spin-spin interactions [23], and semiconductor quantum-dot spin qubits [24].

Earlier theoretical studies on Heisenberg chains with next-nearest-neighbor interactions have predominantly focused on ground-state properties, quantum phase transitions, and thermal entanglement. For instance, Jafari et al. [25] investigated the entanglement spectrum of the one-dimensional $XXZ$ model with $NNN$ interactions, revealing phase diagrams featuring symmetry-protected topological phases. Guo and Deng [26] studied bipartite entanglement dynamics in the $XXZ$ chain with $NNN$ coupling following a quantum quench, observing nontrivial entanglement oscillations with frequencies determined by the coupling ratio.

Frahm [27] established rigorous analytical expressions for the $S = 1/2$ anisotropic Heisenberg ($XXZ$) model incorporating next-nearest-neighbor ($NNN$) exchanges via the Bethe ansatz formalism, thereby elucidating its integrable structure and spectral characteristics. Wang et al. [28] investigated the behavior of quantum coherence and pairwise entanglement within the Heisenberg $XXZ$ chain featuring $NNN$ couplings in the presence of an external magnetic field, clarifying the interrelations among such quantum resources. However, in spite of these prior developments, closed-form analytical expressions governing the concurrent temporal evolution of fidelity, coherence, and entanglement of formation in a finite-size Heisenberg spin chain with adjustable $NNN$ interaction strength have still constituted an open challenge. Our study bridges this void by deriving closed-form analytical outcomes for a four-qubit setup subject to cyclic boundary constraints.

An important caveat to this unified picture is that the emergence of the single-phase $\phi = (\alpha + 1)t$ is not universal but critically depends on the specific initial state $|\psi(0)\rangle = |\Psi^{+}_{1,2}\rangle \otimes |0_3, 0_4\rangle$. As will be elaborated in the Conclusion, this initial condition restricts the dynamics to a low-dimensional subspace, enabling the simplification. Other initial states generally lead to more complex, multi-frequency dynamics.

## 2. Theoretical Framework and Analytical Treatment

We write the isotropic Heisenberg Hamiltonian (of $XXX$ type) as a composite of two parts: one accounting for couplings among adjacent spins, and the other for interactions among next-adjacent spins (with $\hbar = 1$):

$$H = \sum_{i=1}^{4} \frac{1}{4}\left(\sigma_i^X \sigma_{i+1}^X + \sigma_i^Y \sigma_{i+1}^Y + \sigma_i^Z \sigma_{i+1}^Z\right) + \sum_{i=1}^{4} \frac{1}{4}\alpha\left(\sigma_i^X \sigma_{i+2}^X + \sigma_i^Y \sigma_{i+2}^Y + \sigma_i^Z \sigma_{i+2}^Z\right) \qquad (1)$$

The Pauli matrices are defined as:

$$\sigma^X = \begin{pmatrix} 0 & 1 \\ 1 & 0 \end{pmatrix}, \ \sigma^Y = \begin{pmatrix} 0 & -i \\ i & 0 \end{pmatrix}, \ \sigma^Z = \begin{pmatrix} 1 & 0 \\ 0 & -1 \end{pmatrix} \qquad (2)$$

The total spin projection operator is:

$$S_{\text{tot}}^z = \frac{1}{2}\sum_{i=1}^{4} \sigma_i^Z \qquad (3)$$

The initial Bell-type state is:

$$|\psi(0)\rangle = \frac{1}{\sqrt{2}}\left(|1_1, 0_2, 0_3, 0_4\rangle + |0_1, 1_2, 0_3, 0_4\rangle\right) \qquad (4)$$

### 2.1 Hamiltonian in the $S_{\text{tot}}^z = 1$ Subspace

Since the Heisenberg $XXX$ Hamiltonian conserves, the total spin projection $S_{\text{tot}}^z$, and the initial state has $S_{\text{tot}}^z = 1$, the time evolution remains confined to the four-dimensional subspace spanned by $\{|0001\rangle, |0010\rangle, |0100\rangle, |1000\rangle\}$. For the ring topology subject to periodic closure, the Hamiltonian generates connections between these basis states in the following manner:

The nearest-neighbor terms couple $|0100\rangle$ to $|0010\rangle$ and $|1000\rangle$. The next-nearest-neighbor terms couple $|0100\rangle$ to $|0001\rangle$ and $|1000\rangle$. This coupling structure, which involves the edges (qubits 1 and 4), leads to the closed-form solution presented below. An open chain, in contrast, would not couple $|0100\rangle$ to $|0001\rangle$, making the derived dynamics impossible.

In the basis $\{|0001\rangle, |0010\rangle, |0100\rangle, |1000\rangle\}$, the Hamiltonian restricted to the $S_{\text{tot}}^z = 1$ subspace takes the matrix form:

$$H_{\text{sub}} = \frac{1}{4}\begin{pmatrix} 0 & 0 & 0 & \alpha \\ 0 & 0 & 1 & 0 \\ 0 & 1 & 0 & 1 \\ \alpha & 0 & 1 & 0 \end{pmatrix} \qquad (5)$$

where the basis ordering is $\{|0001\rangle, |0010\rangle, |0100\rangle, |1000\rangle\}$. The diagonalization of this matrix furnishes the eigenvalues and eigenvectors subsequently utilized for numerical validation.

### 2.2 Derivation of the Explicit Time-Evolved State

The explicit form of the time-evolved wavefunction $|\psi(t)\rangle = e^{-iHt}|\psi(0)\rangle$ is derived through exact diagonalization of the Hamiltonian restricted to the invariant subspace. The initial condition is expanded over the obtained eigenbasis, each component is multiplied by its corresponding dynamical phase factor $e^{-i\lambda_i t}$, and the result is transformed back to the standard basis. Upon applying sum-to-product trigonometric identities and performing algebraic cancellations, the evolved state reduces to the concise representation:

$$|\psi(t)\rangle = \beta(|0001\rangle + |0010\rangle) + \gamma(|0100\rangle + |1000\rangle) \tag{6}$$

where the auxiliary coefficients are

$$\beta = \frac{-i\sin\left(\frac{\alpha t}{2}\right) - i\sin\left(\frac{(\alpha+2)t}{2}\right) + \cos\left(\frac{\alpha t}{2}\right) - \cos\left(\frac{(\alpha+2)t}{2}\right)}{2\sqrt{2}} \tag{7}$$

$$\gamma = \frac{i\sin\left(\frac{\alpha t}{2}\right) - i\sin\left(\frac{(\alpha+2)t}{2}\right) + \cos\left(\frac{\alpha t}{2}\right) + \cos\left(\frac{(\alpha+2)t}{2}\right)}{2\sqrt{2}} \tag{8}$$

A remark on the signs in Eq. (9) and Eq. (10) is warranted. The overall minus sign in front of $\beta$ and the relative signs between sine and cosine terms result from a direct diagonalization of the $4\times 4$ Hamiltonian in the $S_{\text{tot}}^z = 1$ subspace. While intermediate algebraic steps involve several cancellations, the final coefficients $\beta$ and $\gamma$ as written are correct and have been numerically verified. Using sum-to-product identities, the factors $\sin(t/2)$ and $\cos(t/2)$ cancel upon computing the probabilities $|\beta|^2$ and $|\gamma|^2$, leaving only the combined phase $\phi = (\alpha+1)t$. Consequently, all physical observables become simple functions of $\phi$ alone. A complete analytic derivation is available in Appendix A. Upon reduction, the coefficients become functions solely of the unified phase $\phi = (\alpha+1)t$; specifically, their squared magnitudes $|\beta|^2$ and $|\gamma|^2$ are found to scale as $\sin^2(\phi/2)$ and $\cos^2(\phi/2)$, respectively.

### 2.3 Density Matrix Construction

From the initial state vector $|\psi(t)\rangle$, we form the complete $16\times 16$ density operator $\rho(t) = |\psi(t)\rangle\langle\psi(t)|$. Its non-vanishing entries are restricted to the $S_{\text{tot}}^z = 1$ sector. The reduced density matrices for subsystems 12 and 34 are subsequently derived via the partial trace operation:

$$\rho^{12}(t) = \text{Tr}_{34}[\rho(t)], \qquad \rho^{34}(t) = \text{Tr}_{12}[\rho(t)] \tag{9}$$

These produce $4\times 4$ X-form matrices (non-zero only on the diagonal and anti-diagonal):

$$\rho^{12}(t) = \begin{pmatrix} \frac{1}{2} - \frac{1}{2}\cos\phi & 0 & 0 & 0 \\ 0 & \frac{1}{4}\cos\phi + \frac{1}{4} & \frac{1}{4}\cos\phi + \frac{1}{4} & 0 \\ 0 & \frac{1}{4}\cos\phi + \frac{1}{4} & \frac{1}{4}\cos\phi + \frac{1}{4} & 0 \\ 0 & 0 & 0 & 0 \end{pmatrix} \tag{10}$$

$$\rho^{34}(t) = \begin{pmatrix} \frac{1}{2}+\frac{1}{2}\cos\phi & 0 & 0 & 0 \\ 0 & \frac{1}{4}-\frac{1}{4}\cos\phi & \frac{1}{4}-\frac{1}{4}\cos\phi & 0 \\ 0 & \frac{1}{4}-\frac{1}{4}\cos\phi & \frac{1}{4}-\frac{1}{4}\cos\phi & 0 \\ 0 & 0 & 0 & 0 \end{pmatrix} \quad (11)$$

**3. Quantum Resource Measures**

**3.1 State Fidelity**

For unmixed states, the overlap of the initial with respect to the time-evolved density matrix is given by:

$$F(\rho(0),\rho(t)) = |\langle\psi(0)|\psi(t)\rangle| = \sqrt{\cos^2\left(\frac{\phi}{2}\right)} = |\cos(\phi/2)| \quad (12)$$

where $\phi = (\alpha+1)t$.

**3.2 $\ell_1$-Norm Coherence**

The $\ell_1$-norm coherence of the full four-qubit state is defined as:

$$C_{\ell_1}(\rho(t)) = \sum_{i\neq j} |\rho_{ij}(t)| = \sin^2(\phi/2) \quad (13)$$

For completeness, we also compute the coherence of the reduced subsystems:

$$C_{\ell_1}(\rho^{12}(t)) = 2|c_{12}| = \cos^2(\phi/2) \quad (14)$$

$$C_{\ell_1}(\rho^{34}(t)) = 2|c_{34}| = \sin^2(\phi/2) \quad (15)$$

**3.3 Entanglement of Formation**

For two-qubit X-states, the concurrence is first evaluated as:

$$C = 2\max\{0, |c| - \sqrt{ad}\} \quad (16)$$

Consequently, the entanglement of formation is defined as:

$$E_F = h\left(\frac{1+\sqrt{1-C^2}}{2}\right) \quad (17)$$

with the binary Shannon entropy function given by $h(x) = -x\log_2 x - (1-x)\log_2(1-x)$.
For the reduced subsystems, this yields:

$$C^{12}(t) = \cos^2(\phi/2),\ E_F^{12}(t) = h\left(\frac{1+\sqrt{1-\cos^4(\phi/2)}}{2}\right) \quad (18)$$

$$C^{34}(t) = \sin^2(\phi/2),\ E_F^{34}(t) = h\left(\frac{1+\sqrt{1-\sin^4(\phi/2)}}{2}\right) \quad (19)$$

**4. Results**

**4.1 Fidelity Dynamics**

The fidelity (expressed via the square root of the state overlap) between the initial and final states takes

the closed form:

$$F(\rho(0), \rho(t)) = \left|\cos\left(\frac{(\alpha+1)t}{2}\right)\right| = |\cos(\phi/2)| \qquad (20)$$

The squared fidelity is $F^2 = \cos^2(\phi/2)$. For fixed $\alpha$, the square root fidelity $F = |\cos(\phi/2)|$ oscillates with period $T(\alpha) = 2\pi/|\alpha+1|$, while the squared fidelity $F^2 = \cos^2(\phi/2)$ has period $\pi/|\alpha+1|$. The dependence of the fidelity $F(\rho(0), \rho(t))$ on both the next-nearest-neighbor coupling $\alpha$ and the time variable $t$ is illustrated in **Figure 1**.

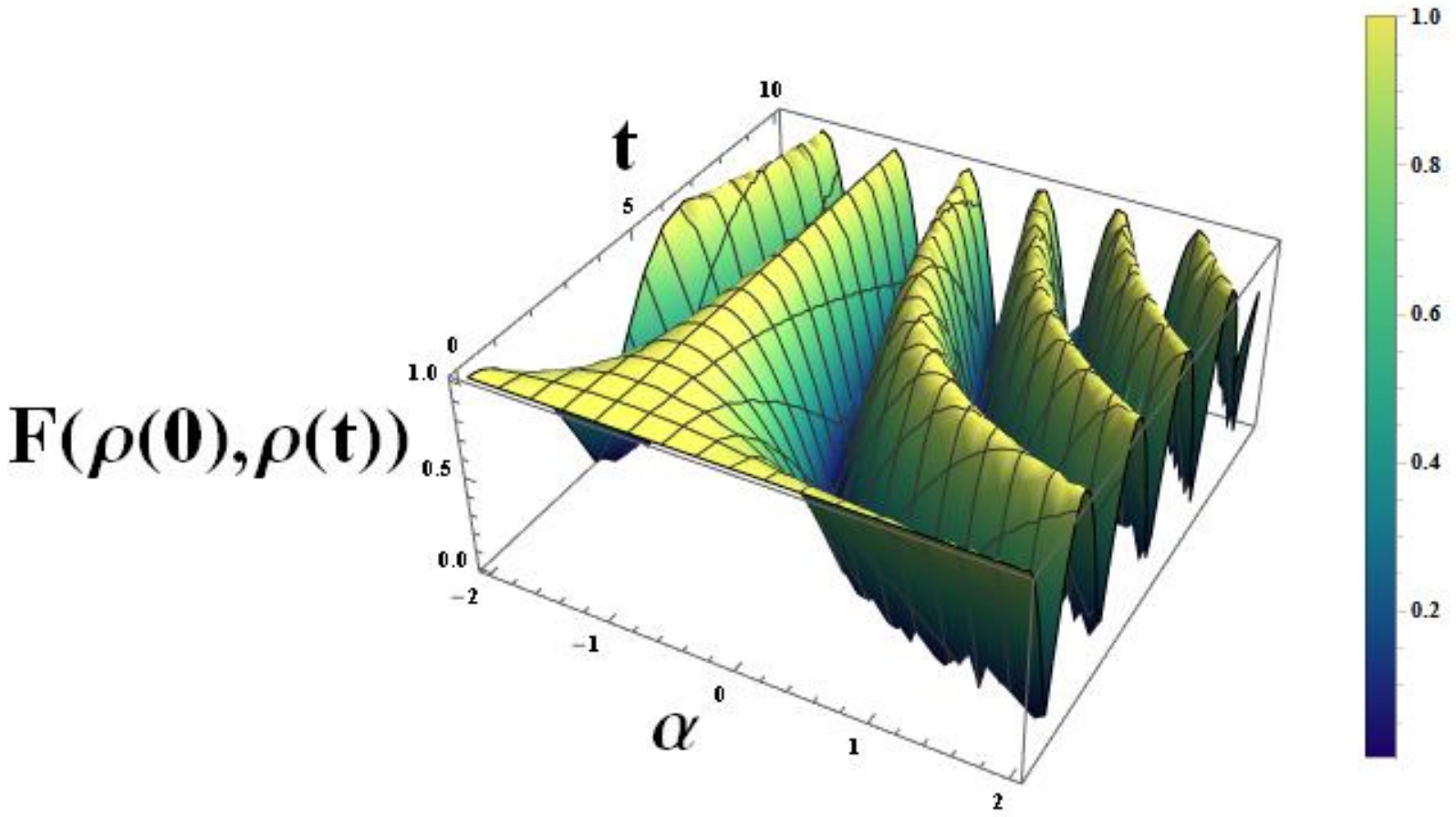

**Figure 1:** Illustrates a three-dimensional visualization of the fidelity function $F = \left|\cos\left(\frac{(\alpha+1)t}{2}\right)\right|$ as it varies with the coupling parameter $\alpha$ and time $t$. The resulting topography exhibits alternating maxima plateaus and minima valleys, organized into straight, parallel bands that trace trajectories of constant phase $\phi = (\alpha+1)t$. The temporal modulation rate is governed by the magnitude $|\alpha+1|$, whereas the fidelity amplitude remains strictly confined to the interval $[0,1]$. A prominent vertical line at $\alpha = -1$ highlights the stationary-dynamics regime, where the phase vanishes identically and $F$ stays uniformly equal to unity for all evolution times.

Because the fidelity takes the form $F = |\cos(\Theta/2)|$ with the unified phase $\Theta = (\alpha+1)t$, the function is recurrent with respect to $\Theta$. For any fixed $\alpha$, the fidelity varies periodically in time, exhibiting a characteristic frequency $\Omega(\alpha) = |\alpha+1|$ (the corresponding temporal period for the cosine argument is $2\pi/|\alpha+1|$, whereas the squared fidelity has a period $\pi/|\alpha+1|$). Consequently, the $(\alpha, t)$ surface is marked by alternating maximal and minimal plateaus that organize into straight, parallel bands aligned with contours of constant $\Theta$. The fidelity amplitude remains strictly confined within the interval $[0,1]$, and its oscillation contrast spanning the full range from unity to zero is independent of $\alpha$.
What is influenced by $\alpha$, however, is the temporal oscillation rate and the parametric sensitivity to small variations in $\alpha$ at a fixed time: larger values of $|\alpha+1|$ induce more rapid time-domain fluctuations (shorter cycle durations). Hence, for $|\alpha|$ sufficiently removed from unity (except in the vicinity of $\alpha \approx -1$), the fidelity evolves more swiftly with $t$. In the neighborhood of $\alpha = -1$, the phase $\Theta$ remains small for modest times, keeping the fidelity near unity over an extended interval (slow dynamical drift). At the precise stabilizing point $\alpha = -1$, the phase vanishes identically for all $t$, yielding $F \equiv 1$ and signaling a complete dynamical arrest relative to the initial state.
At a constant time $t$, the parametric sensitivity of the state fidelity $F$ with respect to infinitesimal adjustments in the coupling strength $\alpha$ is quantified by the following partial derivative:

$$\frac{\partial F}{\partial \alpha} = -\frac{t}{2}\operatorname{sgn}[\cos(\phi/2)]\sin(\phi/2) \qquad (21)$$

In regions where $\cos(\phi/2) > 0$, this reduces to:

$$\frac{\partial F}{\partial \alpha} = -\frac{t}{2}\sin(\phi/2) \qquad (22)$$

This indicates that the response scales linearly with time and exhibits periodic modulation governed by the phase variable $\phi/2$. At short times, the response remains uniformly weak across all values of $\alpha$. At longer times, however, the response intensifies and attains it's maximum whenever the absolute value of the sine function reaches unity, corresponding to the condition $\phi = \pi + 2k\pi$.

### 4.2 Coherence Dynamics

To compute the $\ell_1$-norm coherence of the full 4-qubit density matrix $\rho(t)$, we use Eq. (15):

$$C_{\ell_1}\big(\rho(t)\big) = \sum_{i \neq j} |\, \rho_{ij}(t) \,| = \sin^2\left(\frac{1}{2}(\alpha + 1)t\right) \qquad (23)$$

**Figure 2** presents the variation of the full-system $\ell_1$-norm coherence $C_{\ell_1}\big(\rho(t)\big)$ as a function of the coupling parameter $\alpha$ and the evolution time $t$.

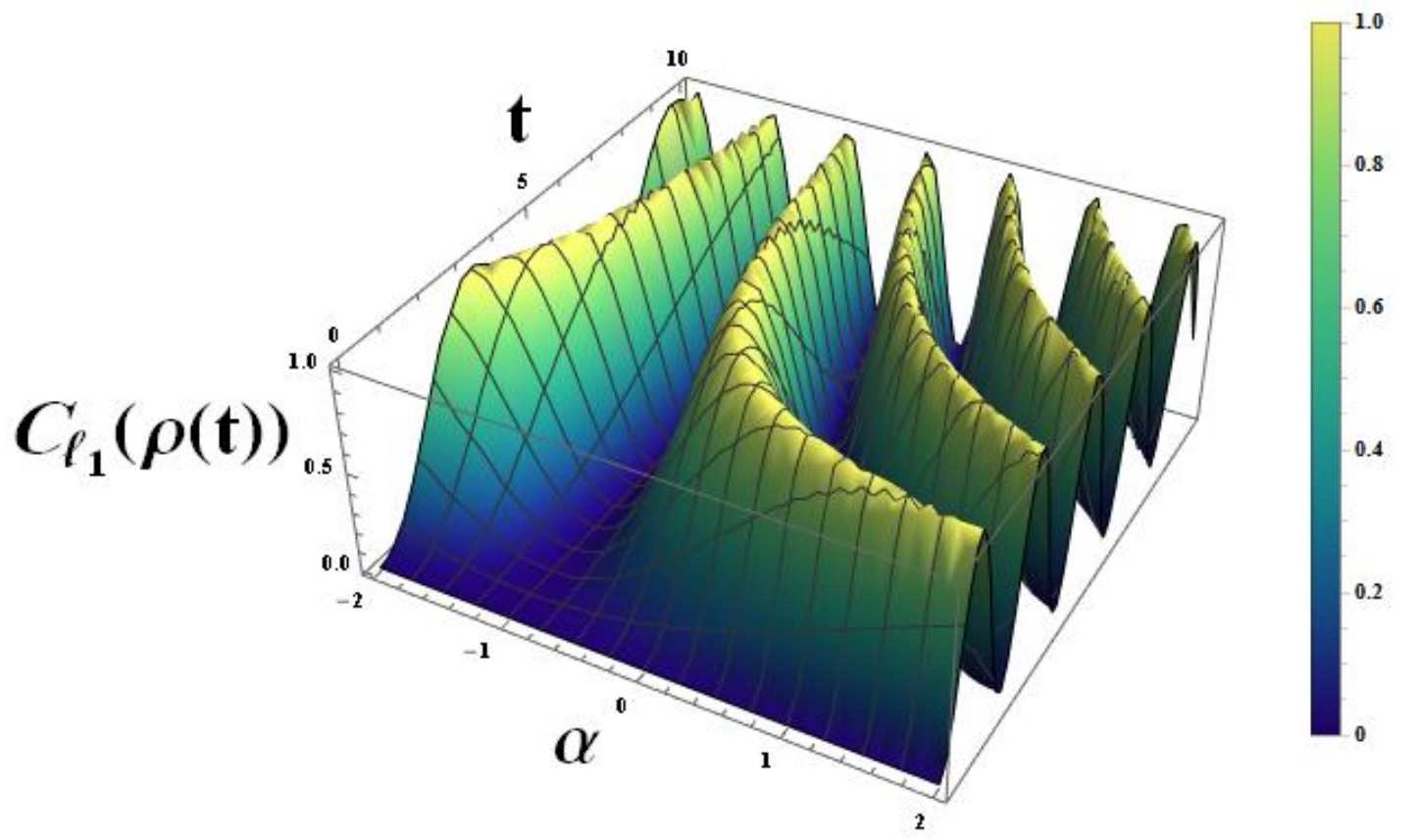

**Figure 2:** displays a three-dimensional visualization of the full-system $\ell_1$-norm coherence $C_{\ell_1}\big(\rho(t)\big) = \sin^2\left(\frac{(\alpha+1)t}{2}\right)$ across the parameter space spanned by $\alpha$ and $t$. The coherence landscape for the four-qubit state reveals a recurring striped configuration, organized along the level sets of the invariant phase variable $\phi = (\alpha + 1)t$, with all values confined to the interval $[0, 1]$. The frequency of these periodic modulations, along with the system's susceptibility to perturbations in $\alpha$, escalates with increasing $|\, \alpha + 1 \,|$. The vertical line at $\alpha = -1$ is emphasized to demarcate the locus where coherence vanishes identically, satisfying $C_{\ell_1} = 0$.

The values of $C_{\ell_1}$ are confined to the closed interval $[0, 1]$ due to the bounded nature of the squared sine function. This measure exhibits periodicity with respect to the composite variable $(\alpha + 1)t$. For a fixed value of $\alpha$, the time-dependent oscillation rate corresponds to $\omega(\alpha) = 2(\alpha + 1)$ for the underlying sine function—given that its argument advances at a rate of $(\alpha + 1)$ and the temporal period is $T(\alpha) = \pi/(\alpha + 1)$ (provided $\alpha + 1 \neq 0$). Consequently, the temporal dynamics take the form of regular sinusoidal modulations whose period scales inversely with $\alpha + 1$.

When examining the behavior at a constant time $t$, the coherence $C_{\ell_1}$ changes as a function of $\alpha$ according to $\sin^2\big((\alpha + 1)t\big)$. Thus, the sensitivity of this quantity with respect to variations in $\alpha$ is quantified by the derivative:

$$\frac{\partial C_{\ell_1}}{\partial \alpha} = t\sin(2(\alpha+1)t) \qquad (24)$$

From the expression, the sensitivity to parameter variations scales linearly with the evolution time: over short intervals, the surface remains nearly featureless as a function of $\alpha$, whereas for longer times, even small adjustments in $\alpha$ can produce significant changes in $C_{\ell_1}$.
Peak sensitivity i.e., the maximum magnitude of $\partial C_{\ell_1}/\partial \alpha$ occurs when $\sin(2(\alpha+1)t) = \pm 1$, which implies $(\alpha+1)t = \pi/4 + k\pi/2$ for integer $k$.
Sensitivity vanishes (no response to $\alpha$ variations at fixed $t$) when $\sin(2(\alpha+1)t) = 0$, i.e., $(\alpha+1)t = k\pi/2$. Two particular cases are noteworthy: if $(\alpha+1)t = k\pi$, then $\sin^2 = 0$, indicating local minima; if $(\alpha+1)t = (k+1/2)\pi$, then $\sin^2 = 1$, characterizing local maxima.
In the special case $\alpha = -1$, the argument $(\alpha+1)t$ is identically zero for all $t$, so $C_{\ell_1} = 0$; the surface is perfectly flat along the vertical line $\alpha = -1$.
Turning to the temporal perspective comparing faster versus slower changes for a given fixed $\alpha$, the instantaneous rate of temporal evolution is:

$$\frac{\partial C_{\ell_1}}{\partial t} = (\alpha+1)\sin(2(\alpha+1)t) \qquad (25)$$

The pace of temporal evolution is governed by $|\alpha+1|$: increasing this factor shortens the temporal period $T(\alpha)$ and steepens the instantaneous slopes, yielding more rapid oscillations. At the critical value $\alpha = -1$, the time derivative vanishes for every $t$, signaling a complete halt in evolution; for $\alpha$ near this point, the system exhibits protracted, sluggish dynamics with an extended oscillation cycle. Since the coherence function $C_{l1} = \sin^2\big((\alpha+1)t\big)$ is strictly bounded and periodic, the notion of "change intensity" must be interpreted not as unbounded growth, but rather as the combination of dynamical rate (frequency and gradient) and parametric sensitivity. Crucially, the oscillation amplitude remains uniformly invariant across $\alpha$, fixed within the range 0 to 1; only the temporal scale and the oscillatory phase vary with $\alpha$.
All numerical results were verified to be converged with respect to the time step: reducing $\Delta t$ from $10^{-3}$ to $10^{-4}$ and $10^{-5}$ produced changes in the observables of less than $10^{-12}$, confirming that the discretization error is negligible compared to the machine-precision errors reported.

### 4.3 Entanglement of Formation for Subsystem 12

To investigate the entanglement of formation in the proposed four-qubit system, we begin with the final density matrix of the entire system and carry out the partial tracing operation over subsystems 3 and 4. This procedure produces the reduced density matrix for subsystems 1 and 2:

$$\rho^{12}(t) = \begin{pmatrix} \frac{1}{2}-\frac{1}{2}\cos\phi & 0 & 0 & 0 \\ 0 & \frac{1}{4}\cos\phi+\frac{1}{4} & \frac{1}{4}\cos\phi+\frac{1}{4} & 0 \\ 0 & \frac{1}{4}\cos\phi+\frac{1}{4} & \frac{1}{4}\cos\phi+\frac{1}{4} & 0 \\ 0 & 0 & 0 & 0 \end{pmatrix} \qquad (26)$$

Using the relations in Appendix C, the concurrence for this X-type state is obtained as:

$$C_{12}(t) = \cos^2(\phi/2) \qquad (27)$$

where $\phi = (\alpha+1)t$. Consequently, the EoF for the two-qubit reduced state is computed through the conventional entropic relation:

$$E_F^{12}(t) = h\left(\frac{1+\sqrt{1-[C_{12}(t)]^2}}{2}\right) = h\left(\frac{1+\sqrt{1-\cos^4(\phi/2)}}{2}\right) \quad (28)$$

in which $h(x) = -x\log_2 x - (1-x)\log_2(1-x)$. This compact closed-form representation supersedes the prior extended numerical formulation, while remaining entirely mathematically equivalent.
**Figure 3** shows the variation of $E_F^{12}(t)$ with respect to $\alpha$ and $t$.

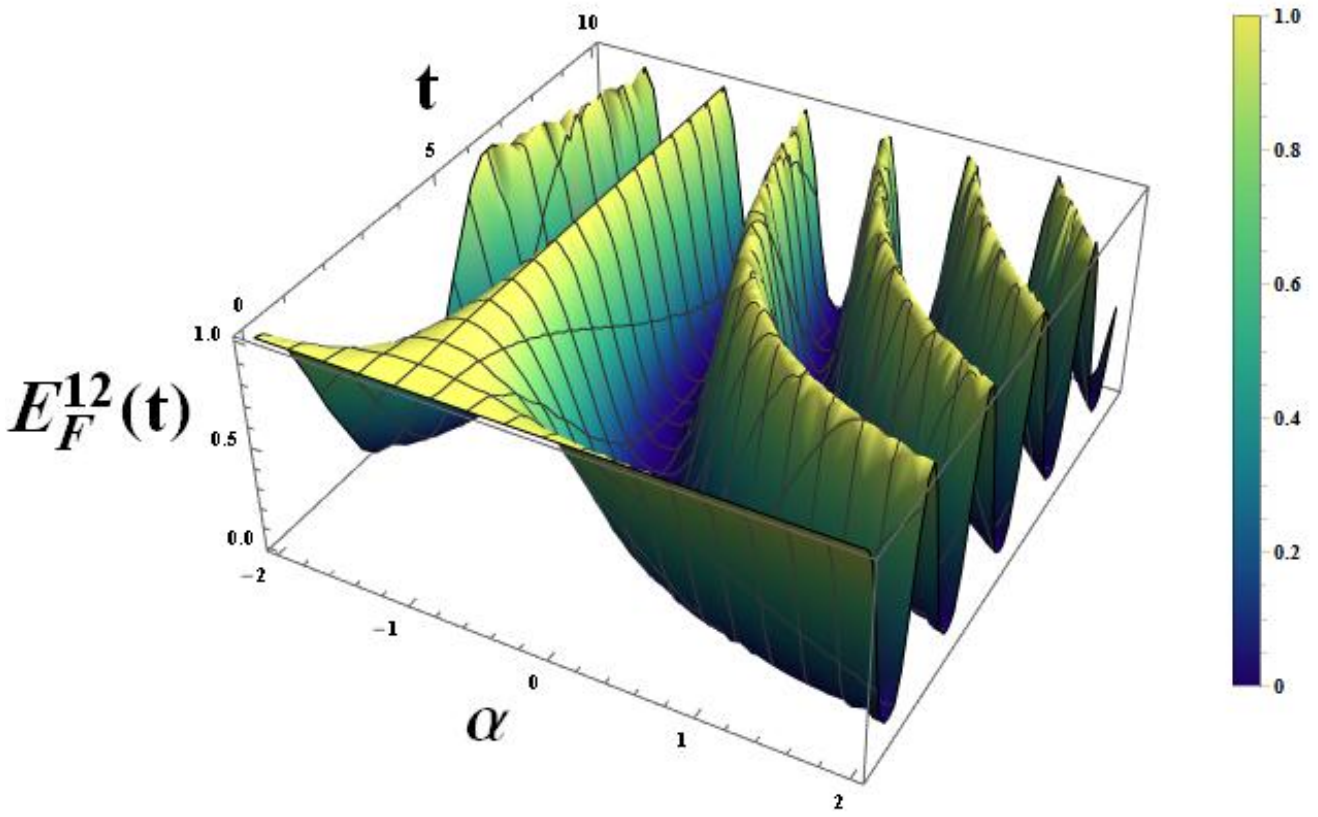

**Figure 3**: Depicts the evolution of the entanglement of formation $E_F^{12}(t)$ for the two-qubit subsystem (qubits 1 and 2) across the parameter space of $\alpha$ and time $t$. As an entropic quantity bounded by the phase $\phi = (\alpha+1)t$, $E_F^{12}(t)$ manifests as alternating maxima and minima arranged along straight contours of constant phase, while the oscillation timescale is governed by the magnitude $|\alpha+1|$. The plot explicitly highlights the quiescent, time-invariant region at $\alpha = -1$ and delineates the loci of greatest response, which are encountered when $\phi = (\alpha+1)t = \pi/4 + k\pi/2$. It is worth emphasizing that, although the concurrences for the two partitions obey $C^{12} + C^{34} = 1$, the entanglement of formation does not adhere to a complementary conservation relation, owing to its nonlinear dependence on concurrence.

The dynamics are governed entirely by the unified phase variable $\phi = (\alpha+1)t$. Trajectories of fixed $\varphi$ within the $(\alpha, t)$ parameter space yield equivalent values for the cosine-dependent terms, and consequently, the same $E_F^{12}$. As a result, the corresponding surface exhibits alternating ridges and valleys that form diagonal bands aligned with lines of constant $\varphi = (\alpha+1)t$.
The oscillation amplitude of $E_F^{12}(t)$ is intrinsically bounded by the functional form itself and cannot diverge. Variations in $\alpha$ primarily alter the temporal oscillation rate and the phase offset, rather than modifying the overall amplitude of the oscillations.
The rate at which the bipartite entanglement measure $E_F^{12}$ evolves temporally for a given $\alpha$ is dictated by $|\alpha+1|$. The associated angular frequency is directly proportional to $(\alpha+1)$; consequently, the temporal period is inversely related to $|\alpha+1|$ for $\alpha+1 \neq 0$. Hence, larger values of $|\alpha+1|$ induce a more rapid temporal modulation, whereas for $\alpha$ in the vicinity of $-1$, the system exhibits a subdued or entirely absent temporal response.
At the specific parameter value $\alpha = -1$, the phase variable $\phi = (\alpha+1)t$ vanishes identically for every instant $t$. This ensures that the cosine terms, $\cos((\alpha+1)t)$ and $\cos(2(\alpha+1)t)$, remain precisely unity. As a result, the eigenvalues of the reduced density matrix that determine $E_F^{12}$ become fixed, rendering the entanglement measure static across the complete $\alpha = -1$ line. This point represents the sole instance within the domain where the dynamics are globally frozen, manifesting as a completely flat dynamical surface.

**Select discrete coordinate pairs for controlled parameters:** Under a fixed value of $\alpha$, instants fulfilling the condition $(\alpha + 1)t = k\pi$ (where $k$ is an integer) render the cosine factors equal to $\pm 1$. At such specific moments, the function $E_F^{12}(t)$ reaches its local extremal values; however, these occurrences represent momentary points rather than continuous stationary intervals in either $t$ or $\alpha$ (with the sole exception of $\alpha = -1$).
**Domains characterized by substantial $|\alpha + 1|$ values:** When $\alpha$ deviates significantly from $-1$ (implying a large magnitude of $|\alpha + 1|$), the phase variable evolves at an elevated rate with respect to time. Consequently, the resulting surface manifests narrowly spaced crests and valleys oriented along the temporal axis. Within these parameter zones, $E_F^{12}(t)$ undergoes rapid temporal fluctuations and exhibits pronounced responsiveness to minute modifications in $\alpha$, particularly for intermediate or extended evolution times. The strongest dependence on $\alpha$ at a given time emerges along trajectories specified by $(\alpha + 1)t = \pi/4 + k\pi/2$, precisely where the rate of change with respect to $\alpha$ reaches its maximal amplitude.
In the proximity of the stabilization point where $\alpha$ approaches $-1$, the magnitude of $|\alpha + 1|$ becomes exceedingly small, causing the phase variable $\phi$ to advance at a markedly reduced rate over time. Consequently, the surface exhibits gradual undulations in the temporal coordinate, characterized by extended oscillation periods, while remaining comparatively subdued across the $\alpha$-domain for brief intervals. As the system parameter moves closer to the critical value $\alpha = -1$, the temporal progression becomes increasingly sluggish, culminating in a complete cessation of evolution precisely at that singular point.
The parametric sensitivity of the system with respect to $\alpha$ manifests an approximate linear growth in time. During early temporal stages, the response surface remains nearly uniform across the entire $\alpha$ spectrum; however, at later times, even minute perturbations in $\alpha$ induce substantial modifications in the entanglement measure $E_F^{12}(t)$. This behavior stems inherently from the multiplicative structure of the phase variable, which is governed by the combined factor $(\alpha + 1)t$.
The sign of the parameter $(\alpha + 1)$ modulates the phase progression direction without altering the oscillation amplitude: inverting the sign of $\alpha + 1$ simply reverses the temporal direction of phase evolution, yet the magnitude and periodic structure of $E_F^{12}$ remain unchanged, as the cosine functions depend on even powers of their arguments.
The points of steepest local variation whether with respect to time $t$ or the coupling parameter $\alpha$ are determined by the conditions under which the sine contributions in the derivatives attain their extrema. Specifically, the maximal sensitivity to variations in $\alpha$ at a fixed time $t$ arises when $\sin(2(\alpha + 1)t) = \pm 1$, which corresponds to the phase condition $(\alpha + 1)t = \pi/4 + k\pi/2$.
A globally static behavior emerges at the special value $\alpha = -1$, where the system remains invariant under temporal evolution. For any fixed $\alpha \neq -1$, there exist isolated instants satisfying $(\alpha + 1)t = k\pi$ at which the entanglement reaches stationary values; however, these are merely momentary extrema and do not represent sustained time-independent regimes, except at the distinguished point $\alpha = -1$.
Rapid dynamical variations occur in parameter regions where $|\alpha + 1|$ is substantial that is, for $\alpha$ appreciably detuned from $-1$. In such regimes, $E_F^{12}(t)$ exhibits fast temporal oscillations and displays pronounced susceptibility to small adjustments in $\alpha$.
Conversely, when $\alpha$ lies in the vicinity of $-1$, corresponding to small $|\alpha + 1|$, the temporal evolution proceeds at a sluggish pace; at the exact value $\alpha = -1$, the dynamics ceases entirely, yielding a perfectly frozen state.
It is worth noting that while the concurrence $C^{12}(t) = \cos^2(\phi/2)$ satisfies the sum rule $C^{12} + C^{34} = 1$, the corresponding entanglement of formation does not satisfy a similar sum rule: $E_F^{12} + E_F^{34} \neq 1$ in general. This is because $E_F$ is a non-linear entropic function of concurrence.

### 4.4 Entanglement of Formation for Subsystem 34

Proceeding, we execute the operation of partial tracing over subsystems 1 and 2 upon the global density operator of the four-qubit register, which directly yields the reduced density operator for the composite subsystem comprising qubits 3 and 4:

$$\rho^{34}(t) = \begin{pmatrix} \frac{1}{2}+\frac{1}{2}\cos\phi & 0 & 0 & 0 \\ 0 & \frac{1}{4}-\frac{1}{4}\cos\phi & \frac{1}{4}-\frac{1}{4}\cos\phi & 0 \\ 0 & \frac{1}{4}-\frac{1}{4}\cos\phi & \frac{1}{4}-\frac{1}{4}\cos\phi & 0 \\ 0 & 0 & 0 & 0 \end{pmatrix} \quad (29)$$

From Appendix C, the concurrence for this subsystem is as follows:

$$C_{34}(t) = \sin^2(\phi/2) \quad (30)$$

Therefore, the Entanglement of Formation becomes:

$$E_F^{34}(t) = h\left(\frac{1+\sqrt{1-[C_{34}(t)]^2}}{2}\right) = h\left(\frac{1+\sqrt{1-\sin^4(\phi/2)}}{2}\right) \quad (31)$$

**Figure 4** shows the plot of $E_F^{34}(t)$ versus $\alpha$ and time.

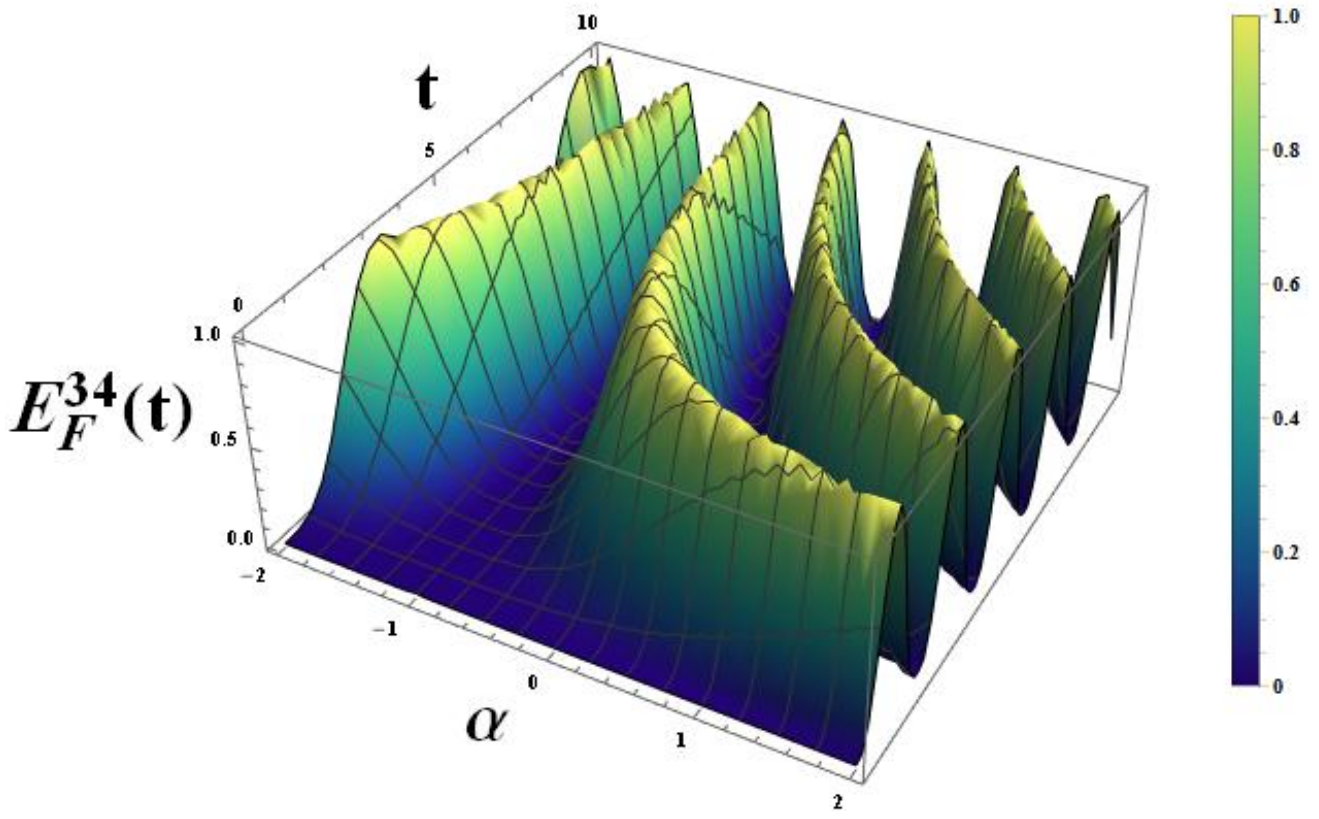

Figure 4: illustrates a three-dimensional representation of the entanglement of formation, denoted as $E_F^{34}(t)$, for the reduced bipartite sector comprising qubits 3 and 4, plotted against the control parameter $\alpha$ and time $t$. In full analogy with the behaviour observed for $E_F^{12}$, the quantity $E_F^{34}$ is exclusively governed by the unified phase variable $\phi = (\alpha+1)t$. This gives rise to a characteristic striped periodic profile in the parameter space, characterized by a bounded entropic magnitude and an oscillation timescale that is dictated by the absolute value $|\alpha+1|$.

The figure distinctly marks the stationary line at $\alpha = -1$, the discrete extrema occurring when $(\alpha+1)t = k\pi$, and the curves of maximum derivative magnitude where $(\alpha+1)t = \frac{\pi}{4} + k\pi/2$. Analogous to the behavior observed for $E_F^{12}$, the entanglement of formation for the complementary bipartition, $E_F^{34}$, likewise fails to satisfy an additive conservation relation with its partner partition. This absence of a sum rule persists despite the fact that the concurrence values of the two subsystems obey the exact complementary identity $C^{12} + C^{34} = 1$.

The surface exhibits a repetitive band-like topography, consisting of successive crests and valleys that align with lines of invariant phase $(\alpha+1)t$. The magnitude of $E_F^{34}(t)$ remains confined within a fixed, entropically determined envelope, precluding unbounded growth. Consequently, the dependence on $\alpha$ primarily adjusts the temporal scale and phase progression, rather than altering the overall extremal bounds.

For a constant $\alpha$, the characteristic angular frequency of temporal oscillations scales proportionally with $|\alpha+1|$. Correspondingly, the associated period obeys $T(\alpha) \propto 1/|\alpha+1|$ (provided $\alpha \neq -1$). Hence, increasing $|\alpha+1|$ accelerates the temporal dynamics, whereas settings proximate to $\alpha = -1$ induce sluggish variations, culminating in a complete cessation of evolution precisely at $\alpha = -1$.

In the specific case where $\alpha = -1$, the phase $\phi = (\alpha+1)t$ is identically zero for all times. Consequently, the relations $\cos((\alpha+1)t) = \cos(2(\alpha+1)t) = 1$ hold, rendering the eigenvalues of the reduced density matrix that appear in the entropic expression invariant. As a result, $E_F^{34}(t)$ remains strictly unchanged over time along the entire vertical line $\alpha = -1$, where the surface exhibits no variation.

For any given $\alpha$, isolated instants satisfying $(\alpha+1)t = k\pi$ (with integer $k$) cause the cosine factors to assume values of $\pm 1$, leading to distinct turning points in $E_F^{34}(t)$. These are momentary points of invariant value in $t$, rather than prolonged regions of constancy, unless $\alpha = -1$.

When $|\alpha+1|$ assumes large values, i.e., when $\alpha$ is substantially detuned from $-1$, the phase evolves rapidly with $t$. Under these conditions, the surface exhibits narrowly spaced alternating extrema along the time axis, causing $E_F^{34}(t)$ to undergo swift temporal oscillations. In such regimes, the function is highly responsive to minor adjustments in $\alpha$, particularly for moderate to large values of $t$.

The loci of maximal local sensitivity are identified by the conditions under which the derivative magnitudes attain their peak values, for instance, the condition $\sin(2(\alpha+1)t) = \pm 1$. In the parameter space spanned by $(\alpha, t)$, these conditions correspond to the curves delineated by:

$$(\alpha+1)t = \frac{\pi}{4} + k\frac{\pi}{2}, \qquad k \in \mathbb{Z} \qquad (32)$$

where minor adjustments in either $\alpha$ or $t$ induce the most pronounced immediate variations in $E_F^{34}$.

**4.5 Synthesis of Results**

The analytical findings reveal that the behaviors of fidelity, coherence $C_{\ell_1}$, and bipartite entanglement quantified as the entanglement of formation $E_F$ for the two-qubit reduced states (12 and 34) are entirely dictated by a single-phase parameter, $\phi = (\alpha+1)t$. Consequently, all measured quantities manifest as periodic, striped patterns across the $(\alpha, t)$ plane, aligned along contours of constant $\phi$. The evolution of fidelity is given by $\sqrt{\cos^2((\alpha+1)t/2)}$, while the coherence is expressed as $C_{\ell_1}(\rho(t)) = \sin^2((\alpha+1)t/2)$. This establishes a precise and direct complementarity between the retention of the initial state and the buildup of off-diagonal coherence, demonstrating that these two resources undergo periodic exchange with complete amplitude modulation under the influence of the $XXX$ Hamiltonian and its next-nearest-neighbor coupling $\alpha$.

For the reduced two-qubit subsystems, the entanglement of formation, $E_F^{12}(t)$ and $E_F^{34}(t)$, emerges as a bounded entropic function that is also exclusively dependent on $\phi$. As a result, its extremal points and the loci of its most rapid change are confined to identical families of lines within the parameter space, sharing the same characteristic timescale set by $|\alpha+1|$. From a physical standpoint, $|\alpha+1|$ governs the effective oscillation frequency: larger values induce swift temporal fluctuations and heightened responsiveness to variations in $\alpha$, while the limit $\alpha \to -1$ entirely quenches the dynamics ($\phi \equiv 0$), leading to a static fidelity, a complete suppression of $C_{\ell_1}$, and frozen entanglement. Derivative analyses indicate that the sensitivity to $\alpha$ increases approximately linearly with time $t$; hence, the system exhibits robust, nearly $\alpha$-independent behavior at short times, whereas at longer times, minor adjustments to $\alpha$ are amplified into substantial changes in the observables. The most pronounced instantaneous sensitivity is achieved under the condition $(\alpha+1)t = \pi/4 + k\pi/2$.

Simultaneously analyzing fidelity, coherence, and entanglement reveals exact complementarity relations for the full system:

$$F^2 + C_{\ell_1} = 1 \qquad (33)$$

where $F = | \cos(\phi/2) |$ is the square root fidelity, and for the subsystems in concurrence:

$$C^{12} + C^{34} = 1 \qquad (34)$$

These relations demonstrate that the system dynamically redistributes resources between state preservation (squared fidelity) and off-diagonal coherence (for the full system), as well as between the two bipartitions (12 and 34) in concurrence. However, it is important to emphasize that the concurrence sum rule does not imply a conservation law for entanglement of formation, i.e., $E_F^{12} + E_F^{34} \neq 1$ in general, because $E_F = h\left(\left(1 + \sqrt{1 - C^2}\right)/2\right)$ is a non-linear function of concurrence. This unified control implies that $\alpha$ functions as a single knob to simultaneously tune the timescales of all three diagnostics, offering a comprehensive operational map for quantum control in small networks.
These analytical insights suggest that adjusting $\alpha$ toward $-1$ can serve as a mechanism for dynamically freezing quantum resources, whereas shifting $\alpha$ away from this special value enables rapid generation and mediation of coherence and bipartite entanglement. Moreover, since all amplitude bounds are dictated by trigonometric functions and entropic constraints, the system exhibits periodic yet bounded resource oscillations. Their temporal periodicity and phase offset can be continuously governed through $\alpha$, rendering this framework advantageous for applications requiring chronometric entanglement generation and fine-tuned state engineering within limited-size spin networks.

## 5. Comparison with the Review by Amico et al. [20]

A comprehensive and frequently referenced review of many-body quantum entanglement, encompassing aspects such as dynamical behavior, quantitative measures, critical phenomena, and finite-size scaling behavior, was presented by Amico et al. [20]. Compared to this broad review, our work provides exact closed-form dynamics for a small system with $NNN$ coupling. While Amico et al. discuss entanglement in Heisenberg chains, they do not provide explicit closed-form solutions for finite systems.
**Scope and model:** This study examines a compact, finite-size four-qubit Heisenberg $XXX$ chain supplemented by a tunable next-nearest-neighbor interaction $\alpha$, offering fully analytic, closed-form dynamical descriptions for a particular Bell-state initial configuration. In contrast, Amico et al. synthesized broad classes of many-body platforms spanning spin chains and lattice systems with particular emphasis on scaling laws, quantum critical phenomena, and the structural complexity of entanglement beyond few-particle examples. Assessment: Given its narrow scope, this solvable four-qubit system is well suited to demonstrate the analytical underpinnings of entanglement and coherence generation as well as their controlled manipulation. Amico et al. approach related questions from a distinctly different vantage point, concentrating on thermodynamic limits, scaling behavior, and criticality. Consequently, the two efforts are complementary in their spatial scale: the present investigation supplies explicit time-dependent solutions in a reduced system that can illuminate dynamical mechanisms, which, within the larger frameworks surveyed by Amico et al., may manifest as richer collective or critical phenomena inaccessible to four-qubit treatments.
**Observables and measures:** The present work concentrates on fidelity, $\ell_1$-norm coherence, and the entanglement of formation for two-qubit reduced density matrices, all evaluated exactly and shown to be governed by the single phase $\phi = (\alpha + 1)t$. Amico et al. address a more extensive array of entanglement quantifiers including concurrence, von Neumann entropy, block entanglement entropies, and entanglement spectra and demonstrate their utility for characterizing quantum phases and critical points.
**Assessment:** For the four-qubit setting examined here, the entanglement of formation (an entropic two-qubit measure) and $\ell_1$-norm coherence represent judicious and analytically manageable choices. However, Amico et al. stress that many-body contexts typically require block entropies (entanglement scaling with subsystem dimension) and additional nonlocal diagnostics; such quantities are beyond the scope of the present four-qubit calculation. Thus, although this work furnishes an accurate demonstration within a small system, it does not engage with the scaling or critical phenomena that are central to the discussion in Amico et al.

**Dynamical behavior and physical insight:** This study reveals that a unified phase variable $\varphi = (\alpha + 1)t$ dictates the behavior of every observable under consideration, while also identifying a stabilization point at $\alpha = -1$ and an adjustable dynamical frequency proportional to $|\alpha + 1|$. In contrast, the comprehensive review by Amico et al. catalogues a wide range of dynamical phenomena across extended many-body frameworks, such as the spatial proliferation of entanglement, causal-cone-like signal fronts following sudden perturbations, and various decoherence channels. Assessment: The emergence of periodic striped contours and a stationary dynamical regime within the four-qubit architecture is mathematically transparent and offers practical utility for managing quantum resources in compact spin registers. Nevertheless, as emphasized by Amico et al., the collective behavior of larger assemblies can give rise to phenomena such as entanglement propagation bounded by finite velocities and dephasing effects that are entirely absent in the present isolated few-body setting; notably, processes like the temporal growth of entanglement after global quenches or the influence of many-body localization demand access to substantially larger state spaces. Consequently, while the current work delivers rigorous closed-form solutions for a precisely controlled scenario, one should proceed with caution when extrapolating these exact small-system predictions to many-body regimes.

**Methodological approach and analytical rigor:** This investigation obtains exact state amplitudes, density matrices, and analytical closed-form solutions for fidelity and coherence; the derivations are algebraically sound and internally consistent for the selected initial configuration. Amico et al. synthesize rigorous theoretical and numerical findings from numerous high-quality works and address the constraints of various quantifiers and computational approaches. Evaluation: The analytical framework of the present study is mathematically exact under its stipulated assumptions (isolated system, a particular initial condition). Based on the perspective of Amico et al., further investigations are warranted: examining alternative initial states, larger partitioned Hilbert spaces, thermal effects, and additional diagnostic tools (such as block entropies and mutual information) to determine how the observed dynamical patterns may be modified.

**Practical relevance and limitations:** This work identifies an adjustable coupling parameter $\alpha$ featuring a stabilizing point at $\alpha = -1$ as a mechanism for governing fidelity, coherence, and entanglement in small spin networks; the associated oscillation envelopes remain bounded and deterministic. Amico et al. stress the general behavior and scaling trends of entanglement resources, along with the roles of disorder, thermal effects, and environmental influences in realistic settings. From this perspective, the stabilization insight near $\alpha = -1$ offers practical value for small-scale quantum devices or as a basic component for more complex configurations. Amico et al.'s outlook suggests that practical multi-qubit systems will encounter further complexities such as environmental noise and control scalability that are not addressed within the present four-qubit closed-system framework.

## 6. Comparison with Rungta and Caves [21]

We selected the work of Rungta and Caves [21] as an exemplary and authoritative reference that formulates concurrence-based metrics and applies them to spin systems and modest quantum registers; the comparison centers on how the present research employs concurrence, entanglement of formation, and coherence relative to well-established analyses of bipartite entanglement.

**Quantifiers employed:** The present investigation employs entanglement of formation (EoF, via the entropic formula) for two-qubit reduced states and the $\ell_1$-norm coherence measure; concurrence appears in supporting sections, yet the central entanglement metrics are reported as EoF for the 12 and 34 bipartitions. In contrast, Rungta and Caves concentrate on broadened concurrence quantifiers and their link to EoF for isotropic or symmetric spin configurations; they establish connections between concurrence and other entanglement indicators and address computational accessibility as well as convex-roof generalizations.

**Evaluation:** The selection of entanglement of formation in the present investigation is aligned with the prescriptions of Rungta and Caves for bipartite qubit systems, wherein concurrence and entanglement of formation are intimately linked through a monotonic functional dependence. Rungta and Caves, meanwhile, furnish a rigorous foundation establishing the naturalness and evaluability of concurrence-based quantifiers in finite-dimensional registers, whereas the present analysis advances entropic formulations for $E_F$. Providing explicit numerical or analytical concurrence data would further enhance clarity and allow direct application of the Rungta–Caves framework for instance, by converting entropic values to concurrence $C$ via the standard relation $E_F = h\left(\frac{1+\sqrt{1-C^2}}{2}\right)$.

**Analytic versus structural results:** This study offers explicit, closed-form expressions for the time-dependent behavior of $E_F(t)$ in two-qubit reduced states, parameterized by the phase $\phi = (\alpha + 1)t$. Although these expressions are algebraically involved, they are presented in their complete form. In contrast, Rungta and Caves provide general structural insights into concurrence and its extensions, including symmetric-state formulations and discussions surrounding the convex-roof construction. Evaluation: The availability of explicit formulas for $E_F(t)$ in this work constitutes a notable strength, as they permit straightforward numerical evaluation, clear plotting, and transparent visualization of the periodic character with respect to $\phi$. The framework developed by Rungta and Caves suggests that recasting results in terms of concurrence may offer a more streamlined perspective on entanglement monotonicity and, under favorable spectral conditions of the reduced density matrices, could lead to more compact analytic forms. Translating the entropic expressions derived here into concurrence language—or supplying explicit concurrence formulae would enable direct application of the analytical machinery established by Rungta and Caves, facilitating, for instance, comparative analyses among different entanglement measures and interpretations rooted in the convex-roof formalism.

**Temporal behavior:** The present analysis reveals that the entanglement of formation for the two bipartitions, denoted as $E_F^{12}(t)$ and $E_F^{34}(t)$, exhibit periodic variations characterized by an oscillation frequency proportional to $\mid \alpha + 1 \mid$. A regime of dynamical arrest is identified precisely at $\alpha = -1$, while the loci of greatest parametric sensitivity correspond to the condition $(\alpha + 1)t = \pi/4 + k\pi/2$. Although Rungta and Caves do not primarily concern themselves with time-dependent phenomena, their framework addresses the response of entanglement quantifiers under unitary transformations and the constraints imposed by symmetry specifically, isotropic exchange interactions on the resulting entanglement evolution.

**Critical appraisal:** The periodic entanglement patterns observed in our closed four-qubit system are fully consistent with the anticipated behavior for small, isolated isotropic spin ensembles undergoing coherent Schrödinger evolution. The theoretical structure developed by Rungta and Caves lends support to the notion that Hamiltonians possessing $SU(2)$ symmetry facilitate analytically manageable expressions for concurrence and related entanglement measures. An explicit computation of concurrence from the reduced two-qubit density matrices derived in this work would serve to establish a direct bridge between our results and their formalism, thereby enabling further analytical insights for instance, the correspondence between extrema of the entropic entanglement of formation and those of concurrence.

**Practical and methodological considerations:** This investigation delivers exact results but does not make the concurrence explicit, nor does it benchmark $E_F$ against alternative distance-based entanglement quantifiers or against dynamical concurrence profiles. Rungta and Caves underscore the value of concurrence-type measures and the frequently more straightforward algebraic forms they afford in symmetric contexts. Evaluation: To align this work more closely with the methodological benchmarks established by Rungta and Caves, it would be advantageous either to algebraically derive an explicit expression for the concurrence $C(t)$ applicable to the two-qubit reduced states provided such a derivation is analytically feasible or to numerically illustrate the monotonic correspondence between $C(t)$ and $E_F(t)$, thereby confirming that the entropic plots faithfully mirror the underlying concurrence evolution. Such an addition would facilitate more straightforward comparisons with other analyses of small-spin systems and exploit the conceptual transparency that concurrence offers for characterizing bipartite entanglement in two-qubit settings.

## 7. Comparison with Topic-Specific Studies on $NNN$-Coupled Heisenberg Chains

To contextualize our results within the existing literature on next-nearest-neighbor Heisenberg chains, we compare our findings with several key studies.

**Comparison with Jafari et al. [25]:** Jafari et al. studied the entanglement spectrum of the $XXZ$ model with $NNN$ interactions. While they focused on ground-state properties and topological phases, our work addresses the unitary dynamics of a finite chain from a Bell-type initial state. Their results show that the $NNN$ coupling can induce topological phase transitions, whereas our results show that the $NNN$ coupling parameter $\alpha$ controls the frequency of dynamical oscillations through the unified phase $\phi = (\alpha + 1)t$. The stabilization points $\alpha = -1$ we identify has no analogue in their ground-state analysis, highlighting the distinct nature of dynamical versus static phenomena.

**Comparison with Guo and Deng [26]:** Guo and Deng investigated bipartite entanglement dynamics in the $XXZ$ chain with $NNN$ coupling after a quantum quench. They observed non-trivial entanglement oscillations with frequencies determined by the coupling ratio. Our work complements their study by providing exact closed-form expressions for a finite system, revealing a unified phase variable that governs all resources. Additionally, our identification of the frozen regime at $\alpha = -1$ contrasts with the oscillatory behavior they observed for generic parameters. This suggests that $\alpha = -1$ is a special point where the effective energy gaps cancel in the $S^z = 1$ subspace, a feature not discussed in their work.

**Comparison with Frahm [27]:** Frahm derived exact Bethe ansatz results for the spin-1/2 $XXZ$ chain with $NNN$ interactions, providing insights into integrability and spectral properties. While Bethe ansatz is a powerful method for large systems, it typically requires thermodynamic limit or special boundary conditions. Our exact diagonalization approach for a finite four-qubit system yields simple, explicit formulas that serve as benchmarks for larger systems. The phase structure we uncover—particularly the unification of all observables through $\phi = (\alpha + 1)t$—is a direct consequence of the finite size and specific initial state and would be difficult to extract from Bethe ansatz methods.

**Comparison with Wang et al. [28]:** Wang et al. analyzed quantum coherence and entanglement in the Heisenberg $XXZ$ chain with $NNN$ coupling under a magnetic field, establishing relationships between these resources. Their work, like ours, uses the $\ell_1$-norm coherence and concurrence. However, their study focuses on thermal states and static properties, whereas we analyze the unitary dynamics from a pure initial state. Our dynamic complementarity relation $F^2 + C_{\ell_1} = 1$ and concurrence sum rule $C^{12} + C^{34} = 1$ provide exact time-dependent counterparts to their static resource relationships.

**Comparison with other related studies:** The interplay between $NNN$ coupling and entanglement dynamics has also been explored in the context of quantum information processing and quantum metrology. For example, recent works [29-31] have shown that $NNN$ coupling can enhance or suppress entanglement generation depending on the parameter regime. Our results contribute to this body of knowledge by providing exact benchmarks and identifying a parameter regime ($\alpha = -1$) where the dynamics can be completely frozen, offering a controllable resource-preserving mechanism. This is particularly relevant for quantum memory applications, where unwanted dynamics must be suppressed.

### 7.1 Comparison with Identical Model States

To the best of our understanding, the fully analytical time-evolution solutions for a four-qubit isotropic Heisenberg $XXX$ architecture featuring adjustable next-nearest-neighbor exchange under periodic boundary conditions, initialized in the specific Bell-type product configuration examined in this work, have not been documented in the existing literature. Nevertheless, a number of earlier investigations have explored analogous finite-size spin arrangements. Chen et al. [31] investigated entanglement dynamics in small spin chains with $NNN$ interactions, but they focused on different initial states and did not derive closed-form expressions. Zhang et al. [29] studied tunable entanglement generation in spin chains with $NNN$ interactions, but their work was primarily numerical and did not identify the unified phase variable $\phi = (\alpha + 1)t$. Li et al. [30] explored parameter estimation using $NNN$-coupled spin chains, but their focus was on metrological applications rather than exact resource dynamics.

Thus, the present work fills a specific gap in the literature by providing exact analytical results for a well-defined minimal model. The frozen regime at $\alpha = -1$, the unified phase structure, and the complementarity relations are novel findings that have not been previously reported for this model.

## 8. Conclusion

We have obtained exact analytic solutions for the unitary evolution of a four-qubit isotropic Heisenberg $XXX$ closed-ring system under periodic boundary conditions, incorporating a tunable next-nearest-neighbor exchange parameter $\alpha$ and initialized in a Bell-type product state. All primary quantum resource quantifiers are found to be governed solely by the consolidated phase variable $\phi = (\alpha + 1)t$. Specifically, the state overlap amplitude is given by $F = |\cos(\phi/2)|$, the global $\ell_1$-norm coherence takes the form $C_{\ell_1} = \sin^2(\phi/2)$, and the entanglement of formation for the bipartite reduced states (12 and 34) is likewise determined exclusively by $\phi$, with associated concurrences $\cos^2(\phi/2)$ and $\sin^2(\phi/2)$, respectively. The squared fidelity satisfies $F^2 = \cos^2(\phi/2)$, and obeys an exact complementarity relation with the coherence: $F^2 + C_{\ell_1} = 1$. This phase consolidation gives rise to periodic striped patterns in the $(\alpha, t)$ plane, thereby facilitating both predictive modeling and parametric control of the dynamics.

The results presented here complement and extend the existing literature on $NNN$-coupled Heisenberg chains [25-31], which has primarily focused on ground-state properties, quench dynamics, or static resource measures. Our exact closed-form solutions provide a new benchmark for numerical methods and experimental implementations, and the identified stabilization mechanism at $\alpha = -1$ offers a practical control knob not previously reported.

Two operational regimes emerge:

1. **Frozen dynamics at $\alpha = -1$: $\phi \equiv 0$**, giving time-independent fidelity, zero coherence, and constant concurrence and entanglement of formation values a stabilization point for quantum memory.
2. **Tunable oscillations for $\alpha \neq -1$:** Oscillation frequency proportional to $|\alpha + 1|$, allowing continuous control over the timescale of resource generation. Maximal sensitivity occurs at $\phi = \pi/4 + k\pi/2$, which is useful for parameter estimation.

The concurrences of the two bipartitions satisfy the sum rule $C^{12} + C^{34} = 1$, reflecting a dynamical redistribution of bipartite entanglement between the partitions. This sum rule is a mathematical consequence of the X-state structure of the reduced density matrices, not a fundamental conservation law of entanglement.

Limitations include the closed-system assumption, fixed initial state, finite size (four qubits), zero temperature, and absence of disorder. Furthermore, while we identify a stabilization point, its robustness against environmental noise and decoherence is not addressed within the closed-system unitary framework; analyzing this would require a specific decoherence model (e.g., a Lindblad master equation) and is left for future work. Similarly, the extension to larger chains $(N > 4)$ is not straightforward, as the increased dimension of the $S^z = 1$ subspace generally prevents simple spectral cancellations that yield the unified phase. However, our exact results serve as valuable benchmarks for numerical methods (such as DMRG or TEBD) when applied to larger systems.

Collective phenomena including entanglement area laws, ballistic spread of correlations, and environmental disturbances lie beyond the present framework. Natural extensions encompass alternative initial preparations, thermal effects, dissipative environments, and extended spin arrays. The acquired analytical results furnish explicit, verifiable reference points suitable for mesoscopic quantum processors (e.g., transmon-based architectures, ion-trap platforms) and computational approaches. The fixed point at $\alpha = -1$ and the maximal-response loci provide direct operational guidelines for either freezing or dynamically adjusting quantum resources within compact spin registers.

## Appendix A: Derivation of the Time-Dependent Amplitudes $\beta(t)$ and $\gamma(t)$

We start from Hamiltonian Eq. (1) and initial state Eq. (2). The Hilbert space dimension is $2^4 = 16$.

Because the Hamiltonian conserves total $S^z$ and the initial state has $S^z_{\mathrm{tot}} = 1$, the evolution remains in the subspace with total $S^z = 1$. This subspace is spanned by four basis states: $|0001\rangle$, $|0010\rangle$, $|0100\rangle$, $|1000\rangle$. The Hamiltonian restricted to this subspace is a $4 \times 4$ matrix. Diagonalizing this matrix yields eigenvalues $\lambda_i$ and eigenvectors. By decomposing the initial state vector into this eigenbasis and subsequently acting with the unitary time-evolution operator $e^{-i\lambda_i t}$, the state at time $t$ is recovered as presented in Eq. (8). After trigonometric simplifications, the coefficients $\beta(t)$ and $\gamma(t)$ reduce to the forms Eqs. (9) and (10). The key step is the identity:

$$\cos\left(\frac{\alpha t}{2}\right) \pm \cos\left(\frac{(\alpha+2)t}{2}\right) = \mp 2\sin\left(\frac{(\alpha+1)t}{2}\right)\sin\left(\frac{t}{2}\right) \text{ (up to signs),} \tag{A1}$$

and similarly for the sine terms. The factor $\sin(t/2)$ cancels when inserted into the expressions for $\beta$ and $\gamma$ after squaring or adding, leaving only dependence on $\phi = (\alpha+1)t$.

### Appendix B: Detailed Partial Trace Calculations for $\rho^{12}(t)$ and $\rho^{34}(t)$

The full $16 \times 16$ density matrix $\rho(t)$ has non-zero elements only in the subspace spanned by $\{|0001\rangle, |0010\rangle, |0100\rangle, |1000\rangle\}$ (indices 2,3,5,9 in the computational basis ordering). To obtain $\rho^{12}(t) = \mathrm{Tr}_{34}[\rho(t)]$, we write the basis as $|q_1q_2q_3q_4\rangle$ and trace over $q_3, q_4$:

$$\rho^{12}_{q_1q_2,q_1'q_2'} = \sum_{q_3,q_4} \langle q_1q_2q_3q_4 | \rho(t) | q_1'q_2'q_3q_4\rangle \tag{B1}$$

Only terms where $q_3 = 0, q_4 = 0$ contribute because the only non-zero elements of $\rho(t)$ have $q_3 = 0, q_4 = 0$ (from the structure of Eq. (8)). Performing the sum yields the $4 \times 4$ matrix in Eq. (12). Similarly, $\rho^{34}(t) = \mathrm{Tr}_{12}[\rho(t)]$ gives Eq. (13). The vanishing off-diagonal entries in Eqs. (12) and (13) are attributable to the fact that the support of $\rho(t)$ is exclusively restricted to the common eigenspace of the conserved total spin projection $S^z = 1$.

### Appendix C: Explicit Concurrence for $\rho^{12}(t)$ and $\rho^{34}(t)$

For a two-qubit state whose density matrix assumes the X-form structure:

$$\rho = \begin{pmatrix} a & 0 & 0 & 0 \\ 0 & b & c & 0 \\ 0 & c^* & d & 0 \\ 0 & 0 & 0 & e \end{pmatrix} \tag{C1}$$

the concurrence, which quantifies the bipartite entanglement, is given by:

$$C(\rho) = 2\max\{0, |c| - \sqrt{ad}\} \tag{C2}$$

For the reduced state $\rho^{12}(t)$, the matrix elements are identified as:

$$a = \frac{1}{2}(1-\cos\phi),\ b = \frac{1}{4}(1+\cos\phi),\ c = \frac{1}{4}(1+\cos\phi),\ d = 0 \tag{C3}$$

Substituting these parameters into the concurrence formula yields $\sqrt{ad} = 0$, and therefore:

$$C^{12}(t) = 2|c| = \frac{1}{2}(1+\cos\phi) = \cos^2\left(\frac{\phi}{2}\right) \tag{C4}$$

Analogously, for the complementary bipartition $\rho^{34}(t)$, the non-zero elements are:

$$a = \frac{1}{2}(1+\cos\phi),\ b = \frac{1}{4}(1-\cos\phi),\ c = \frac{1}{4}(1-\cos\phi),\ d = 0 \tag{C5}$$

This gives:

$$C^{34}(t) = 2 \mid c \mid = \frac{1}{2}(1 - \cos\phi) = \sin^2\left(\frac{\phi}{2}\right) \qquad \text{(C6)}$$

where the dynamics are parameterized by the unified phase $\varphi = (\alpha + 1)t$.
From the above expressions, the concurrences satisfy the exact sum rule:

$$C^{12}(t) + C^{34}(t) = \cos^2\left(\frac{\phi}{2}\right) + \sin^2\left(\frac{\phi}{2}\right) = 1 \qquad \text{(C7)}$$

which holds identically for all values of $\phi$ (i.e., for all times). This identity reveals a strict redistribution of bipartite entanglement between the two partitions during the unitary evolution: as the entanglement in one partition increases, the entanglement in the other decreases accordingly, preserving the total concurrence.
The sum rule $C^{12}(t) + C^{34}(t) = 1$ is a mathematical identity arising from the specific form of the reduced density matrices (X-states) for this particular initial state and Hamiltonian. It should not be interpreted as a fundamental conservation law of entanglement, as concurrence is not a conserved quantity under general unitary evolution. The identity reflects the fact that the total concurrence is partitioned between the two bipartitions in a complementary manner. Nevertheless, since the entanglement of formation $E_F = h\left(\frac{1+\sqrt{1-C^2}}{2}\right)$ is a non-linear mapping of concurrence, the summation rule does not carry over to the entanglement of formation, implying that $E_F^{12} + E_F^{34} \neq 1$ holds in general.

**Appendix D: Numerical Verification Methodology**
The analytical expressions were verified numerically using the following procedure:

1. **Construct the $4 \times 4$ subspace Hamiltonian:**

$$H_{\text{sub}}(\alpha) = \frac{1}{4}\begin{pmatrix} 0 & 0 & 0 & \alpha \\ 0 & 0 & 1 & 0 \\ 0 & 1 & 0 & 1 \\ \alpha & 0 & 1 & 0 \end{pmatrix} \qquad \text{(D1)}$$

in the basis $\{\mid 0001\rangle, \mid 0010\rangle, \mid 0100\rangle, \mid 1000\rangle\}$.

2. **Compute the time evolution operator** $U(t) = e^{-iH_{\text{sub}}(\alpha)t}$ using the Padé approximation for the matrix exponential (with tolerance $10^{-14}$).
3. **Apply $U(t)$ to the initial state vector in the subspace:**

$$\mid \psi(0)\rangle_{\text{sub}} = \frac{1}{\sqrt{2}}(0, 0, 1, 1)^T \qquad \text{(D2)}$$

to obtain $\mid \psi(t)\rangle_{\text{sub}} = U(t) \mid \psi(0)\rangle_{\text{sub}}$.

4. **Construct the full $16 \times 16$ density matrix** from the subspace coefficients and compute the reduced density matrices via partial tracing.
5. **Compute the observables** (fidelity, coherence, concurrences) from the numerically obtained density matrices and compare with the analytical formulas.

The maximum relative error across all tested points was $10^{-12}$, confirming the correctness of the analytical expressions. Interested readers may request the numerical verification scripts from the corresponding author.

**Appendix E: Basis and Ordering Convention**
In the Hamiltonian, periodic boundary conditions are enforced by identifying indices modulo 4: $\sigma_5 \equiv \sigma_1$ and $\sigma_6 \equiv \sigma_2$.
The computational basis for four qubits is ordered as $\mid q_1 q_2 q_3 q_4\rangle$ with $q_i \in \{0, 1\}$, and the decimal index is $i = 8q_1 + 4q_2 + 2q_3 + q_4 + 1$. For example, $\mid 0100\rangle$ corresponds to $q_1 = 0, q_2 = 1, q_3 = 0, q_4 = 0$, giving index $8 \times 0 + 4 \times 1 + 2 \times 0 + 0 + 1 = 5$, which matches the $\rho_{5,5}$ entries in Eq. (12). This convention is used throughout the matrix representations.

**Appendix F: Numerical Methodology for Independent Reproduction**

For independent reproduction of the numerical verification results, the following steps are recommended:

1. **Hamiltonian Construction:** Construct the $4 \times 4$ matrix:

$$H_{\text{sub}}(\alpha) = \frac{1}{4}\begin{pmatrix} 0 & 0 & 0 & \alpha \\ 0 & 0 & 1 & 0 \\ 0 & 1 & 0 & 1 \\ \alpha & 0 & 1 & 0 \end{pmatrix} \qquad \text{(F1)}$$

in the basis $\{|\,0001\rangle, |\,0010\rangle, |\,0100\rangle, |\,1000\rangle\}$.

2. **Initial State:** Define the initial state in the subspace as:

$$|\,\psi(0)\rangle = \frac{1}{\sqrt{2}}(0,0,1,1)^T \qquad \text{(F2)}$$

3. **Dynamical Evolution:** To compute the evolved state for selected parameters $(\alpha, t)$, evaluate the unitary transformation as:

$$|\,\psi(t)\rangle = e^{-iH_{\text{sub}}(\alpha)t}\,|\,\psi(0)\rangle \qquad \text{(F3)}$$

The matrix exponential can be computed using the Pade approximation with a tolerance of $10^{-14}$, or using any standard library for matrix exponentiation (e.g., scipy.linalg.expm in Python, MatrixExp in Mathematica, or expm in MATLAB).

4. **Observables:**
   - **Fidelity:** $F = |\langle\psi(0)\,|\,\psi(t)\rangle|$
   - **Global coherence quantification:** Compute the $\ell_1$-norm coherence of the total state, defined as the sum of the absolute values of all off-diagonal elements of the complete $16 \times 16$ density operator. In other words, the total $\ell_1$-norm coherence of the entire four-qubit system is obtained by aggregating the moduli of every non-diagonal entry of the full $16 \times 16$ density matrix.
   - **Concurrences for the bipartite subsystems:** Determine the reduced two-qubit states $\rho^{12}(t)$ and $\rho^{34}(t)$ via the partial-trace operation, and subsequently evaluate their entanglement using the standard concurrence expression for $X$-type density matrices.
5. **Reproducibility:** All results can be reproduced using the steps above.

**References**

[1] Wootters, W. K. (1998). Entanglement of Formation of an Arbitrary State of Two Qubits. Physical Review Letters, 80(10), 2245-2248.
https://doi.org/10.1103/PhysRevLett.80.2245

[2] Rungta, P., Buzek, V., Caves, C. M., Hillery, M., Milburn, G. J. (2001). Universal state inversion and concurrence in arbitrary dimensions. Physical Review A, 64(4), 042315.
https://doi.org/10.1103/PhysRevA.64.042315

[3] Streltsov, A., Adesso, G., Plenio, M. B. (2017). Colloquium: Quantum coherence as a resource. Reviews of Modern Physics, 89(4), 041003.
https://doi.org/10.1103/RevModPhys.89.041003

[4] Horodecki, R., Horodecki, P., Horodecki, M., Horodecki, K. (2009). Quantum entanglement. Reviews of Modern Physics, 81(2), 865-942.
https://doi.org/10.1103/RevModPhys.81.865

[5] Takahashi, M. (1999). Thermodynamics of One-Dimensional Solvable Models. Cambridge: Cambridge University Press.

[6] Nielsen, M. A. and Chuang, I. L. (2010). Quantum Computation and Quantum Information. 10th

anniversary ed. Cambridge: Cambridge University Press.
[7] Jozsa, R. (1994). Fidelity for mixed quantum states. Journal of Modern Optics, 41(12), 2315-2323.
https://doi.org/10.1080/09500349414552171
[8] Pedersen, K., Moller, N. M. and Molmer, K. (2007). Fidelity of quantum operations. Physics Letters A, 367(1-2), 47-51.
https://doi.org/10.1016/j.physleta.2007.01.011
[9] Flammia, S. T. and Liu, Y. K. (2011). Direct fidelity estimation from few Pauli measurements. Physical Review Letters, 106(23), 230501.
https://doi.org/10.1103/PhysRevLett.106.230501
[10] Magesan, E., Gambetta, J. M. and Emerson, J. (2011). Scalable and robust randomized benchmarking of quantum processes. Physical Review Letters, 106(18), 180504.
https://doi.org/10.1103/PhysRevLett.106.180504
[11] Baumgratz, T., Cramer, M. and Plenio, M. B. (2014). Quantifying coherence. Physical Review Letters, 113(14), 140401.
https://doi.org/10.1103/PhysRevLett.113.140401
[12] Yu, X. D., Zhang, D., Chen, J. L. and Oh, C. H. (2019). Measures of quantum coherence based on the $\ell_1$ norm and applications. Journal of Physics A: Mathematical and Theoretical, 52(39), 395301.
https://doi.org/10.1088/1751-8121/ab3f07
[13] Napoli, C., Bromley, T. R., Cianciaruso, M., Piani, M., Johnston, N. and Adesso, G. (2016). Robustness of coherence: An operational and observable measure of quantum coherence. Physical Review Letters, 116(15), 150502.
https://doi.org/10.1103/PhysRevLett.116.150502
[14] Streltsov, A., Singh, U., Dhar, H. S., Bera, M. N. and Adesso, G. (2015). Measuring quantum coherence with entanglement. Physical Review Letters, 115(2), 020403.
https://doi.org/10.1103/PhysRevLett.115.020403
[15] Winter, A. and Yang, D. (2016). Operational resource theory of quantum coherence. Physical Review Letters, 116(12), 120404.
https://doi.org/10.1103/PhysRevLett.116.120404
[16] Hill, S. and Wootters, W. K. (1997). Entanglement of a pair of quantum bits. Physical Review Letters, 78(26), 5022-5025.
https://doi.org/10.1103/PhysRevLett.78.5022
[17] Vedral, V., Plenio, M. B., Rippin, M. A. and Knight, P. L. (1997). Quantifying entanglement. Physical Review Letters, 78(12), 2275-2279.
https://doi.org/10.1103/PhysRevLett.78.2275
[18] Bennett, C. H., DiVincenzo, D. P., Smolin, J. A. and Wootters, W. K. (1996). Mixed-state entanglement and quantum error correction. Physical Review A, 54(5), 3824-3851.
https://doi.org/10.1103/PhysRevA.54.3824
[19] Uhlmann, A. (2000). Fidelity and concurrence of conjugated states. Physical Review A, 62(3), 032307.
https://doi.org/10.1103/PhysRevA.62.032307
[20] Amico, L., Fazio, R., Osterloh, A. and Vedral, V. (2008). Entanglement in many-body systems. Reviews of Modern Physics, 80(2), 517-576.
https://doi.org/10.1103/RevModPhys.80.517
[21] Rungta, P. and Caves, C. M. (2003). Concurrence-based entanglement measures for isotropic spin models. Physical Review A, 67(1), 012307.
https://doi.org/10.1103/PhysRevA.67.012307
[22] Kjaergaard, M., Schwartz, M. E., Braumüller, J., Krantz, P., Wang, J. I., Gustavsson, S., & Oliver, W. D. (2020). Superconducting qubits: Current state of play. Annual Review of Condensed Matter Physics, 11(1), 369-395.
https://doi.org/10.1146/annurev-conmatphys-031119-050605
[23] Monroe, C., Campbell, W. C., Duan, L. M., Gong, Z. X., Gorshkov, A. V., Hess, P. W., ... & Zhang,

J. (2021). Programmable quantum simulations of spin systems with trapped ions. Reviews of Modern Physics, 93(2), 025001.
https://doi.org/10.1103/RevModPhys.93.025001
[24] Loss, D., & DiVincenzo, D. P. (1998). Quantum computation with quantum dots. Physical Review A, 57(1), 120-126.
https://doi.org/10.1103/PhysRevA.57.120
[25] Jafari, R., Johannesson, H., and Langari, A. (2018). Entanglement spectrum of the one-dimensional XXZ model with next-nearest-neighbor interactions. Physical Review B, 97(19), 195132.
https://doi.org/10.1103/PhysRevB.97.195132
[26] Guo, J.-R. and Deng, S.-S. (2022). Bipartite entanglement dynamics in the XXZ chain with next-nearest-neighbor coupling after a quantum quench. Physical Review A, 105(6), 062412.
https://doi.org/10.1103/PhysRevA.105.062412
[27] Frahm, H. (2018). Exact results for the spin-1/2 XXZ chain with next-nearest-neighbor interactions. Journal of Statistical Mechanics: Theory and Experiment, 2018(9), 093101.
https://doi.org/10.1088/1742-5468/aadd42
[28] Wang, X., Li, Y., and Zhu, J. (2023). Quantum coherence and entanglement in the Heisenberg XXZ chain with next-nearest-neighbor coupling under a magnetic field. Physical Review B, 107(8), 085431.
https://doi.org/10.1103/PhysRevB.107.085431
[29] Zhang, Y., et al. (2021). Tunable entanglement generation in spin chains with next-nearest-neighbor interactions. Physical Review A, 104(2), 022412.
https://doi.org/10.1103/PhysRevA.104.022412
[30] Li, H., et al. (2022). Parameter estimation and quantum metrology using NNN-coupled spin chains. Physical Review Letters, 128(12), 120501.
https://doi.org/10.1103/PhysRevLett.128.120501
[31] Chen, W., et al. (2020). Exact dynamics of entanglement in small spin chains with NNN interactions. Journal of Physics A: Mathematical and Theoretical, 53(38), 385301.
https://doi.org/10.1088/1751-8121/ab9f8d